# Power-law behaviors from the two-variable Langevin equation：Ito's and Stratonovich's Fokker-Planck equations

Guo Ran, Du Jiulin[*]

*Department of Physics, School of Science, Tianjin University, Tianjin 300072, China*

**Abstract**

We study power-law behaviors produced from the stochastically dynamical system governed by the well-known two-variable Langevin equations. The stationary solutions of the corresponding Ito's, Stratonovich's and the Zwanzig's (the backward Ito's) Fokker-Planck equations are solved under a new fluctuation-dissipation relation, which are presented in a unified form of the power-law distributions with a power index containing two parameter $\kappa$ and $\sigma$, where $\kappa$ measures a distance away from the thermal equilibrium and $\sigma$ distinguishes the above three forms of the Fokker-Planck equations. The numerical calculations show that the Ito's, the Stratonovich's and the Zwanzig's form of the power-law distributions are all exactly the stationary solutions based on the two-variable Langevin equations.

## 1. Introduction

Non-exponential and/or power-law distributions have been found frequently in some complex systems away from equilibrium, such as the kappa-distributions measured by the observations of the solar wind and space plasmas [1-6], and those $\alpha$-distributions noted in physics, chemistry and elsewhere like $P(E) \sim E^{-\alpha}$ with an index $\alpha >0$ [2, 7-11]. Some of the power-law distributions in the complex systems may be modeled by the $q$-distributions within nonextensive statistical mechanics [12]. Because these power-law behaviors are non-Maxwellian, which have gone beyond the realm governed by the traditional statistical mechanics, the investigations on the physical mechanism to generate these power-law distributions and their stochastically dynamical origins underlying some complex systems away from equilibrium are becoming important endeavors for understanding the nature in many different processes of physical, chemical, biological, technical and their inter-disciplinary fields. For these purposes, we need to analyze the long-time stationary behaviors of the stochastic differential equations associated with some complex dynamical processes, including Langevin equations and the associated Fokker-Planck (F-P) equations and, of course, some Boltzmann equation approaches. But, usually it is difficult to solve a general multi-variable F-P equation for a complex system. In fact, not much has been known in general about the long time stationary-state behaviors of an arbitrary F-P equation. Thus the works on power-law distributions have basically focused on some of single-variable Langevin equations and/or the associated single-variable F-P equations [2, 13-22].

---
[*] Email address: jiulindu@yahoo.com.cn

In the past many years, one has been trying a variety of ways and means to find the dynamical origin of power-law distributions. However, this long-standing problem is not easily solved because, in addition to theoretical complexity, it is often linked with the complex systems so called in many different processes of physical, chemical, biological, technical and their inter-disciplinary fields. The theoretical works on the topic using one-variable Langevin equations and/or the associated one-variable F-P equations to study power-law distributions might include, for instance, a velocity-space F-P equation in the superthermal radiation field plasma [2], a linear F-P equation and the Ito-Langevin equation [15], a velocity-space linear Langevin equation and the $\beta$–fluctuation in the F-P equation [18], an anomalous diffusion of momentum space in an optical lattice [19], a multiplicative noise using a momentum space F-P equation [20], a nonlinear F-P equation in an overdamped motion [21], and recently the theory of power-law reaction dynamics [22] etc. Additionally, the readers might find the examples of kinetic theories that yield power-law distributions, such as the q-generalized Boltzmann equation [23] and the astrophysical systems with self-gravitating long-range interactions [24]. Most recently, one has generally studied the well-known two-variable Langevin equations for the system with a space/velocity-dependent inhomogeneous friction and noise and has exactly generated power-law distributions of many different forms from the associated two-variable Zwanzig's F-P equation [25]. These power-law distributions have been obtained by finding out a new fluctuation-dissipation relation (FDR) for the nonequilibrium dynamical system.

In this work, we follow the idea of Ref.[25] to study the following questions that have naturally raised in the current focus of concern. In the traditional Ito's and Stratonovich's F-P equation, can there be the stationary solutions of power-law distributions? What are the mathematical forms of power-law distributions? Are they different from those forms based on the Zwanzig's F-P equation? Which one can be the exact stationary solution of the two-variable Langevin equations governing the nonequilibrium dynamical system? For these questions, in sec.2, we study the power-law distributions based on the associated Ito's and Stratonovich's F-P equations with the two-variable Langevin equations. And then, in sec.3, we present the numerical analyses of the power-law behaviors based on the Langevin equations. Finally in sec.4, we give the conclusions.

## 2. The analytic solutions of Ito's and Stratonovich's F-P equations

The two-variable Langevin equation is well-known for position and momentum $(x, p)$, modeling the Brownian motion of a particle moving in a potential field $V(x)$. If the medium is inhomogeneous and the noise is multiplicative (space/velocity dependent), one can write the Langevin equation as

$$\frac{dx}{dt} = \frac{p}{m}, \quad \frac{dp}{dt} = -\frac{dV(x)}{dx} - \gamma(x,p)p + \eta(x,p,t). \tag{1}$$

As usual, the noise is Gaussian and satisfys the zero average and the delta-correlated in time $t$,

$$\langle \eta(x,p,t) \rangle = 0, \quad \langle \eta(x,p,t)\eta(x,p,t') \rangle = 2D(x,p)\delta(t-t'), \tag{2}$$

where $m$ is the particle's mass. But, the friction coefficient $\gamma(x,p)$ and the diffusion coefficient $D(x, p)$ are now both a function of position and momentum and follow a generalized fluctuation-dissipation relation [25] given by

$$D = \gamma m \beta^{-1}(1-\kappa\beta E), \qquad (3)$$

where $\beta=1/kT$ with $T$ temperature and $k$ Boltzmann constant, $E$ is the particle's energy, $E = V(x)+p^2/2m$, and $\kappa \neq 0$ is a parameter which measures a distance away from the thermal equilibrium. In the case of $\kappa = 0$, the standard FDR $D = m\gamma\beta^{-1}$ is recovered from Eq.(3). Some examples of anomalous diffusions have implied that the diffusion coefficient may not only depend on the kinetic energy (the square of momentum or velocity) [2, 14, 16, 19], but also depend on the potential energy [26-29]. From the recent experimental study of anomalous diffusion in driven-dissipative dusty plasma, we can get a strong insight into the dependence of diffusion coefficient on the interaction potential [30, 31]. The situations that friction coefficient depends on the kinetic energy (the square of momentum or velocity) can be found in the discussions such as canonical-dissipative systems [32], nonlinear Brownian motion [33] or in the book [34] etc.

Under the new FDR, the corresponding F-P equation in Zwanzig's rule (also called backward Ito's rule [36]) exactly has a stationary-state solution of power-law distributions [25],

$$\rho_s(E) = Z^{-1}(1-\kappa\beta E)_+^{1/\kappa}, \qquad (4)$$

where the normalization constant is $Z = \iint dxdp\,(1-\kappa\beta E)_+^{1/\kappa}$ and $(y)_+ = y$ for $y>0$ and zero otherwise. In the limit $\kappa \to 0$, it recovers Maxwell-Boltzmann distribution.

Having given a Langevin equation, one will be faced with the question how to interpret them with the noises. Different interpretations lead to different drift terms appeared in the corresponding F-P equation. For the conventional Stratonovich's and Ito's rules and the backward Ito's rule (i.e. Zwanzig's rule), readers can see the Appendix A in Ref.[36] for clear interpretations and/or can also see books such as Ref. [35] for Zwanzig's rule, and Refs. [37] [38] for Stratonovich's and Ito's rules. Now we can study the power-law behaviors from the two-variable Langevin equation and the corresponding Ito's and Stratonovich's F-P equations. If $\rho(x,p,t)$ is the probability distribution function of variables $(x, p)$ at time $t$, corresponding to Eqs.(1)

and (2), the Stratonovich's F-P equation [36] is written as

$$\frac{\partial \rho}{\partial t} = -\frac{p}{m}\frac{\partial \rho}{\partial x} + \frac{\partial}{\partial p}\left[\frac{dV}{dx} + \gamma p - D^{\frac{1}{2}}\frac{\partial D^{\frac{1}{2}}}{\partial p}\right]\rho + \frac{\partial^2 (D\rho)}{\partial p^2}, \quad (5)$$

which can be expressed equivalently as

$$\frac{\partial \rho}{\partial t} = -\frac{p}{m}\frac{\partial \rho}{\partial x} + \frac{\partial}{\partial p}\left[\frac{dV}{dx} + \gamma p + \frac{1}{2}\frac{\partial D}{\partial p}\right]\rho + \frac{\partial}{\partial p}D\frac{\partial}{\partial p}\rho. \quad (6)$$

(Here and hereafter, for economization of space, the independent variables of the functions are left out in the equations). On the other hand, the corresponding Ito's F-P equation [36] is written as

$$\frac{\partial \rho}{\partial t} = -\frac{p}{m}\frac{\partial \rho}{\partial x} + \frac{\partial}{\partial p}\left(\frac{dV}{dx} + \gamma p\right)\rho + \frac{\partial^2}{\partial p^2}(D\rho), \quad (7)$$

which can be expressed as

$$\frac{\partial \rho}{\partial t} = -\frac{p}{m}\frac{\partial \rho}{\partial x} + \frac{\partial}{\partial p}\left[\frac{dV}{dx} + \gamma p + \frac{\partial D}{\partial p}\right]\rho + \frac{\partial}{\partial p}D\frac{\partial}{\partial p}\rho. \quad (8)$$

Thus, introducing a parameter $\sigma$, we can write Eq.(6), Eq.(8) and the Zwanzig's F-P equation [25] in a unified form,

$$\frac{\partial \rho}{\partial t} = -\frac{p}{m}\frac{\partial \rho}{\partial x} + \frac{\partial}{\partial p}\left[\frac{dV}{dx} + \gamma p + \sigma\frac{\partial D}{\partial p}\right]\rho + \frac{\partial}{\partial p}D\frac{\partial}{\partial p}\rho. \quad (9)$$

Eq.(9) is Stratonovich's F-P equation if $\sigma = 1/2$, it is Ito's F-P equation if $\sigma = 1$, and it is Zwanzig's F-P equation if $\sigma = 0$. Now we can follow the idea of Ref.[25] to discuss power-law behaviors of the two-variable Langevin equation from the F-P equation Eq.(9). If $\rho_s(x,p)$ is a stationary-state solution of Eq.(9), it satisfies

$$-\frac{p}{m}\frac{\partial \rho_s}{\partial x} + \frac{\partial}{\partial p}\left[\frac{dV}{dx} + \gamma p + \sigma\frac{\partial D}{\partial p}\right]\rho_s + \frac{\partial}{\partial p}D\frac{\partial}{\partial p}\rho_s = 0. \quad (10)$$

Usually, when the friction coefficient and the diffusion coefficient are both constants and they are related with each other by the standard FDR, the solution is Maxwell-Boltzmann (M-B) distribution, being a function of the energy. Now in the case that the friction coefficient and diffusion coefficient are both a function of the variables $x$ and $p$, as a reasonable consideration, we can seek a stationary-state solution of Eq.(10) in the form of $\rho_s = \rho_s(E)$ with the energy $E \equiv V(x)+p^2/2m$. And then, we have

$$\frac{\partial \rho_s}{\partial x} = \frac{dV}{dx}\frac{d\rho_s}{dE}, \quad \frac{\partial \rho_s}{\partial p} = \frac{p}{m}\frac{d\rho_s}{dE}, \tag{11}$$

and Eq.(10) becomes

$$\frac{\partial}{\partial p}\left[\gamma p + \sigma\frac{\partial D}{\partial p} + \frac{p}{m}D\frac{d}{dE}\right]\rho_s = 0. \tag{12}$$

After integrating this equation with respect to $p$, we find

$$\left[\gamma p + \sigma\frac{\partial D}{\partial p} + \frac{p}{m}D\frac{d}{dE}\right]\rho_s = C(x), \tag{13}$$

where $C(x)$ is an "integral constant". In order to solve this differential equation and accordingly study the power-law behaviors of its solutions, we use the generalized FDR condition, Eq.(3), under which the power-law distributions are generated. The following three possible cases are taken into consideration:

(a) If the friction coefficient is a constant, but the diffusion coefficient is a function of the energy, the FDR is

$$D(E) = m\gamma\beta^{-1}(1-\kappa\beta E). \tag{14}$$

In this case, Eq.(13) becomes

$$p\left[\gamma(1-\sigma\kappa E) + \frac{D}{m}\frac{d}{dE}\right]\rho_s = C(x). \tag{15}$$

Because $C(x)$ is independent of $p$, considering $E \equiv V(x)+p^2/2m$ and then letting $p=0$ in Eq.(15), we can determine $C(x)=0$ exactly. And then the solution is solved directly in a unified form as

$$\rho_s = Z_{\sigma,\kappa}^{-1}(1-\kappa\beta E)^{\frac{1-\sigma\kappa}{\kappa}}, \tag{16}$$

where $Z_{\sigma,\kappa} = \iint dxdp(1-\kappa\beta E)^{\frac{1-\sigma\kappa}{\kappa}}$ is the normalization factor. Thus, the Ito's, Stratonovich's and Zwanzig's forms of the solutions of Eq.(13) are not identical to each other. They are all the power-law distribution with a power containing two parameters $\kappa$ and $\sigma$, and they all recover the M-B distribution when we take the limit $\kappa \to 0$. Early in 1968, from the spectral observations within the plasma sheet, Vasyliũnas first wrote an empirical energy $\kappa'$-distribution for the electrons at high energies [39],

$$\rho(E) \sim \left[1 + \frac{1}{\kappa'}\frac{E}{E_0}\right]^{-(\kappa'-1)}, \tag{17}$$

where $E_0 = kT$ is the most probable energy. It is worth to noting that, only if we make such a parameter transformation, $-1/\kappa' \to \kappa$, this empirical expression is exactly the Ito's form of Eq.(16) at $\sigma = 1$, namely:

$$\rho_s = Z_{1,\kappa}^{-1}(1-\kappa\beta E)^{\frac{1-\kappa}{\kappa}}. \tag{18}$$

(b) If the diffusion coefficient is constant, but the friction coefficient is a function of the energy, the FDR is

$$\gamma(E) = D\beta m^{-1}(1-\kappa\beta E)^{-1}. \tag{19}$$

In this case, the term containing $\sigma$ on the left hand side of Eq.(13) disappears, and thus the solutions of Eq.(13) are identical for the Ito's, Stratonovich's and Zwanzig's form. It is easily shown that the stationary solution is the well-known power-law $\kappa$-distribution [25]:

$$\rho_s = Z_\kappa^{-1}(1-\kappa\beta E)^{\frac{1}{\kappa}}, \tag{20}$$

where $Z_\kappa = \iint dxdp(1-\kappa\beta E)^{\frac{1}{\kappa}}$ is the normalization factor.

(c) More generally, if the friction and diffusion coefficients are both a function of the energy, the FDR is

$$D(E) = \gamma(E)m\beta^{-1}(1-\kappa\beta E). \tag{21}$$

In this case, Eq.(13) can be written by

$$p\left[\gamma + \frac{\sigma}{m}\frac{dD}{dE} + \frac{D}{m}\frac{d}{dE}\right]\rho_s = C(x), \tag{22}$$

Because $C(x)$ is independent of $p$, letting $p=0$ in Eq.(22), we can determine $C(x)=0$. And then, we obtain

$$\rho_s = Z_{\sigma,\kappa}^{-1}\exp\left(-m\int\frac{\gamma}{D}dE\right)\exp\left(-\sigma\int\frac{1}{D}\frac{dD}{dE}dE\right). \tag{23}$$

Using the FDR Eq.(13), the solution of this equation with the following power-law behavior is unified by

$$\rho_s = Z_{\sigma,\kappa}^{-1} (1-\kappa\beta E)^{\frac{1}{\kappa}} D^{-\sigma}. \tag{24}$$

where $Z_{\sigma,\kappa} = \iint (1-\kappa\beta E)^{\frac{1}{\kappa}} D^{-\sigma} dxdp$ is the normalization factor. It is clear that the power-law behaviors of the Ito's form ($\sigma=1$) and the Stratonovich's form ($\sigma=1/2$) of the solution are different from the Zwanzig's form ($\sigma=0$). They are the power-law $\kappa$-distribution but have a factor that depends on the diffusion coefficient in such a rule as $\sim D^{-\sigma}$. This is a very special nature of the Ito's and the Stratonovich's solution in this case.

## 3. Numerical analyses

From the well-known two-variable Langevin equations and the corresponding Ito's, Stratonovich's and Zwanzig's F-P equations, we have derived the stationary solutions of power-law distributions under the condition of the generalized FDR. Generally speaking, the Ito's, the Stratonovich's and the Zwanzig's solutions of the power-law distributions are not identical to each other, which are written by a unified form having a power containing two parameters $\kappa$ and $\sigma$, with $\sigma=1$ for the Ito's solution, $\sigma=1/2$ for the Stratonovich's solution and $\sigma=0$ for the Zwanzig's solution. In order to examine the Ito's, the Stratonovich's and the Zwanzig's form of the power-law distributions so as to show which one may be more accurately the solution of the Langevin equations, we employ the numerical method provided in Ref.[40, 41] to calculate the stationary solutions of the two-variable Langevin equations for the Ito's, Stratonovich's and Zwanzig's rules, respectively. And then we make use of the kernel density estimation (KDE) to obtain the probability distribution function (PDF).

In the numerical calculations, the friction coefficient is set constant ($\gamma=1$) and the diffusion coefficient is set energy-dependent (i.e., the case (a)). The other parameters are set as follows: the Brownian particle mass $m=1$, the inverse temperature $\beta=1$ and the potential function $V(x)=x^2$. The calculations are made for $\kappa=-0.2, -0.5$, respectively. We have simulated 1,000,000 particles in the numerical calculations, and starting from the position $x(t=0)=0$ and the momentum $p(t=0)=0$, we obtained the positions and the momentums of these particles approaching to a stationary state. After getting the solutions of the Langevin equations which can be regard as a set of samples in the view of statistics, the method of KDE has been used to produce the PDF from these samples. In the KDE, the Silverman's rule was used to determine bandwidth and the Gaussian kernel was used to estimate the PDF. Finally the numerical solutions of the PDF were found, which can be compared with the analytic solutions based on Eq.(16). The numerical solutions of the PDF and the analytic solutions given by Eq.(16) for Ito's, Stratonovich's and Zwanzig's rules have been illustrated in Fig. 1, Fig.2 and Fig.3, respectively. It is shown that the Ito's, the Stratonovich's or the Zwanzig's solutions of the power-law distributions, Eq.(16), are all excellently fitting with the numerical solutions from the two-variable Langevin equations.

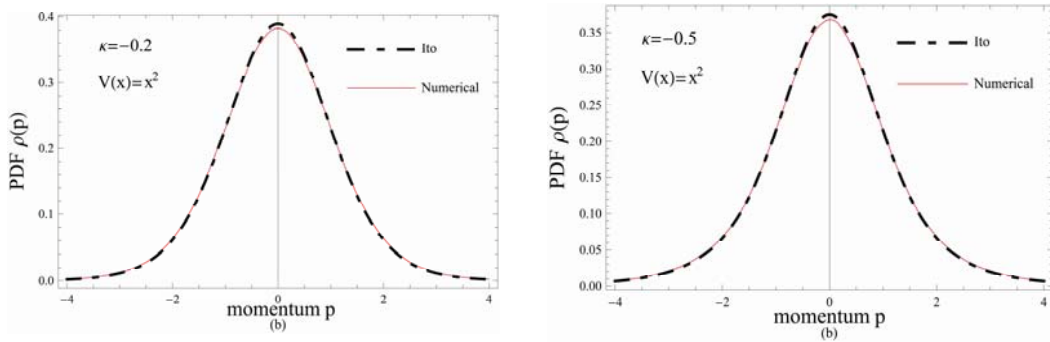

Fig.1. The Ito's solutions of power-law distributions in Eq.(16) and the numerical solutions from the Langevin equations.

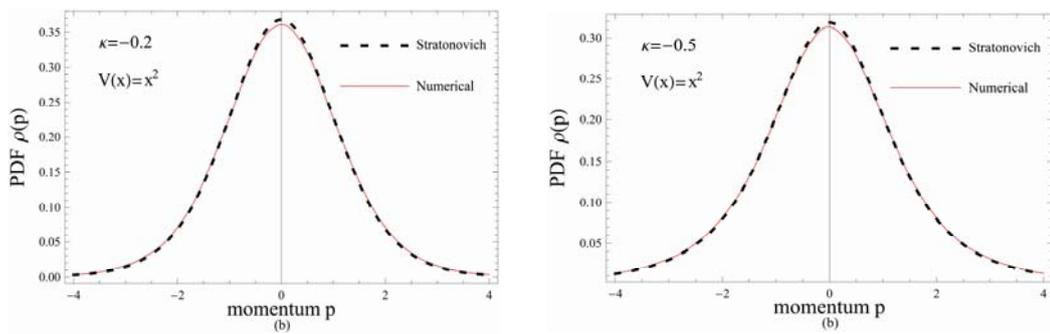

Fig.2. The Stratonovich's solutions of power-law distributions in Eq.(16) and the numerical solutions from the Langevin equations.

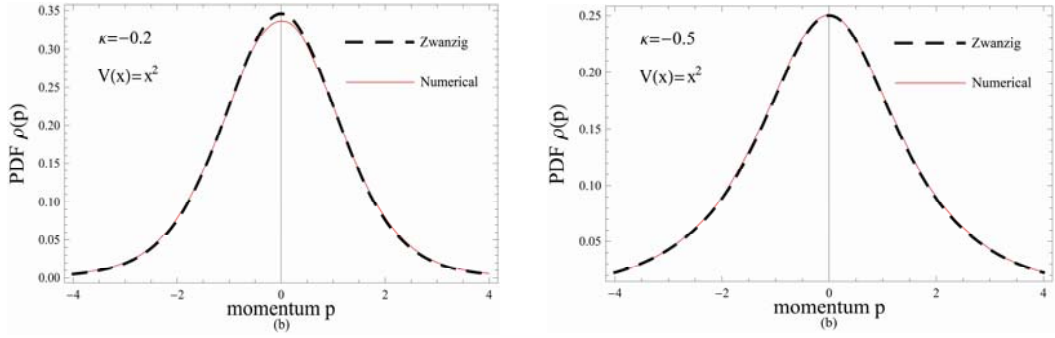

Fig.3. The Zwanzig's solutions of power-law distributions in Eq.(16) and the numerical solutions from the Langevin equations.

**4. Conclusions**

In conclusion, from the well-known two-variable Langevin equations and the corresponding Ito's, Stratonovich's and Zwanzig's F-P equations, we have exactly derived the stationary solutions of power-law distributions under the condition of the generalized FDR. These power-law distributions are obtained in a unified form for the solution of the stationary Ito's, Stratonovich's and Zwanzig's F-P equations, with a power containing two parameters $\kappa$ and $\sigma$, where $\kappa \neq 0$ measures a distance away from the thermal equilibrium. The solution is the Ito's form of power-law distributions if we take $\sigma=1$, is the Stratonovich's form if we take $\sigma=1/2$, and is the Zwanzig's form if we take $\sigma=0$.

The power-law behaviors of the stationary solutions have been discussed in the three cases for the generalized FDR Eq.(3). In the case (a) that the friction coefficient is a constant, but the diffusion coefficient is a function of the energy, the stationary distributions are presented by Eq.(16). The Ito's form, the Stratonovich's form and the Zwanzig's form of the solutions all have an obvious power-law behavior, with a power index containing the parameter $\sigma$ different from each other. Especially, the Ito form of Eq.(16) is found to be exactly the kappa-distribution observed for the electrons at high energies within the plasma sheet in 1968 [39]. In the case (b) that the diffusion coefficient is constant, but the friction coefficient is a function of the energy, the stationary distributions are the power-law $\kappa$-distribution [25] given by Eq.(20), and they are identical for the Ito's, the Stratonovich's and the Zwanzig's form. In the case (c) that the friction and diffusion coefficients are both a function of the energy, the stationary distributions are presented by Eq.(24) for the Ito's, the Stratonovich's and the Zwanzig's form. They all have the power-law behavior of the $\kappa$-distribution, but the Ito's and the Stratonovich's form of the solutions are with a factor that depends on the diffusion coefficient by the power-law $D^{-\sigma}$.

In our numerical calculations, we have shown that the Ito's, the Stratonovich's or the Zwanzig's form of the power-law distributions are all exactly the stationary solutions generated from the two-variable Langevin equations under the new FDR.

**Acknowledgements**

This work is supported by the National Natural Science Foundation of China under Grant No.11175128, and by the Higher School Specialized Research Fund for Doctoral Program under Grant No. 20110032110058.